\documentclass[aps,showpacs,twocolumn,twoside,pre]{revtex4}
\usepackage{graphicx}
\usepackage{amsmath}
\usepackage{amssymb}
\usepackage{mathtext}

\begin{document}

\newcommand {\be}{\begin{equation}}
\newcommand {\ee}[1] {\label{#1} \end{equation}}
\newcommand {\e} {\varepsilon}
\newcommand {\ph} {\varphi}
\newcommand {\lla} {\left\langle}
\newcommand {\rra} {\right\rangle}
\newcommand {\la} {\langle}
\newcommand {\ra} {\rangle}

\title{Coherence versus reliability of stochastic oscillators with delayed feedback}

\author{Denis S.\ Goldobin}
\affiliation{Department of Theoretical Physics, Perm State University,
        15 Bukireva str., 614990, Perm, Russia}

\begin{abstract}
For noisy self-sustained oscillators, both reliability, stability
of a response to a noisy driving, and coherence understood in the
sense of constancy of oscillation frequency belong to the main
characteristics. Though the both characteristics and techniques
for controlling them received great attention of researchers,
owing to their importance for neurons, lasers, clocks, electric
generators, {\em etc.}, these characteristics were previously
considered separately. In this paper, strong quantitative relation
between coherence and reliability is revealed for a limit cycle
oscillator subject to a weak noisy driving and a linear delayed
feedback, a convectional control tool. Analytical findings are
verified and enriched with a numerical simulation for the Van der
Pol--Duffing oscillator.
\end{abstract}

\maketitle

Recently, robustness of response of a limit cycle oscillator to a
noisy driving have attracted considerable attention of both
experimentalists and
theoreticians~\cite{neurons,Uchida-Mcallister-Roy-2004,
Pikovsky-1984, reliability-collected,
Teramae-Tanaka-2004-Goldobin-Pikovsky-2005,
Goldobin-Pikovsky-2005b, Goldobin-Pikovsky-2006,
Nakao-Arai-Kawamura-2007, impulse_noise,telegraph_noise}. In
different fields of science, related phenomena appear under
different names. In neurophysiology the reliability property of
spiking neurons, which manifests itself as a coincidence of
responses of a single neuron to a repeated noisy input of a
prerecorded waveform, attracts great attention~\cite{neurons}. In
recent experiments with a noise-driven Nd:YAG (neodymium-doped
yttrium aluminum garnet) laser~\cite{Uchida-Mcallister-Roy-2004},
a similar property has been referred as consistency. From the
theoretical viewpoint, reliability and consistency are
manifestations of the synchronization of uncoupled nonlinear
oscillators receiving identical noisy
driving~\cite{Pikovsky-1984,reliability-collected,
Teramae-Tanaka-2004-Goldobin-Pikovsky-2005,
Goldobin-Pikovsky-2005b, Goldobin-Pikovsky-2006,
Nakao-Arai-Kawamura-2007, impulse_noise,telegraph_noise}.

Quantitatively, stability of response, reliability, is
characterized by the largest Lyapunov exponent (LE). For smooth
limit cycle oscillators LE is
negative~\cite{Pikovsky-1984,Teramae-Tanaka-2004-Goldobin-Pikovsky-2005,telegraph_noise}
meaning that the system is reliable. However, a large noise may
lead to a positive LE
(\cite{Pikovsky-1984,Schimansky-Geier-Herzel-1993,Goldobin-Pikovsky-2005b,telegraph_noise};
and even antireliability for neuron-like systems in a ``classic''
experimental set-up has been
forecasted~\cite{Goldobin-Pikovsky-2006}).

However, for some oscillatory systems not only the response
stability is important, but also the coherence, {\it i.e.}, the
constancy of the oscillation frequency, which is measured by the
diffusion constant of the oscillation phase. The coherence
determines the precision of clocks (including biological
ones~\cite{biol_clocks}), the quality of electric generators,
susceptibility of an oscillatory system to external
driving~\cite{Goldobin-Rosenblum-Pikovsky-2003}, and
predisposition to synchronization; a laser radiation should be
coherent when one needs to focus the beam or redirect it without
angular divergence; {\it etc}. In
Ref.\,\cite{Goldobin-Rosenblum-Pikovsky-2003} (followed by
methodologically closely related
Ref.\,\cite{Pawlik-Pikovsky-2006-Tukhlina_etal-2008}) the
extremely efficient technique for controlling the coherence by a
weak delayed feedback has been proposed and theoretically analyzed
(a successful experimental implementation of this technique for a
laser in chaotic regime has been reported in
Ref.\,\cite{Boccaletti_etal-2004}). Remarkably, due to the
time-shift symmetry a noiseless limit cycle system is neutrally
stable and remains such in the presence of a delayed feedback. But
in the presence of noise, the delayed feedback utilized for
controlling the coherence may considerably affect the response
stability.

Noteworthy, in the presence of both delay and noise (or
irregularities), the process is not Markovian anymore; therefore,
one may not apply such well-elaborated tools as the conventional
Fokker--Planck equation, and {\it ad hoc} statistical methods are
employed for studies
\cite{ad_hoc-collected,Niebur_etal-1991,Masoller-2002,Frank-collected}.
This paper presents both analytical and numerical results on the
reliability of noisy-driven limit cycle oscillators subject to
delayed feedback control, suggesting an effective mean for
controlling the reliability. Analysis of these results in the
context of controlling coherence reveals strong quantitative
relations [Eq.\,({\ref{e12}})] between the reliability and the
coherence. The disclosed fact, that a high reliability occurs for
a weak coherence, and {\it vice versa} the weaker reliability the
higher coherence, imposes important limitations on implementation
of this conventional control technique. Imperfect cases are also
discussed.

In order to demonstrate numerically the relationship between
coherence and reliability, a simulation for noisy Van der Pol
oscillator
\begin{eqnarray}
&\ddot{x}-\mu(1-x^2)\dot{x}+x=k[\dot{x}(t-\tau)-\dot{x}(t)]+\e\,\xi(t)\,,&
\label{vdp}\\
&\la\xi(t_1)\xi(t_2)\ra=2\,\delta(t_1-t_2)\,,\quad\la\xi\ra=0&\nonumber
\end{eqnarray}
\noindent
has been performed. Here $\mu$ describes closeness to the Hopf
bifurcation point, $k$ and $\tau$ are the feedback strength and
delay time, respectively, $\e$ is the noise amplitude, $\xi(t)$ is
normalized white Gaussian noise. In the presence of noise the
oscillation phase $\ph=-\arctan(x/\dot{x})$ diffuses according to
 $\lla(\ph(t)-\la\ph(t)\ra)^2\rra\propto D\,t$.
Diffusion constant $D$ quantifies the coherence of oscillations.

\begin{figure}[!t]
{\sf
\begin{tabular}{lc}
(a)&\hspace{-6mm}\includegraphics[width=0.435\textwidth]%
 {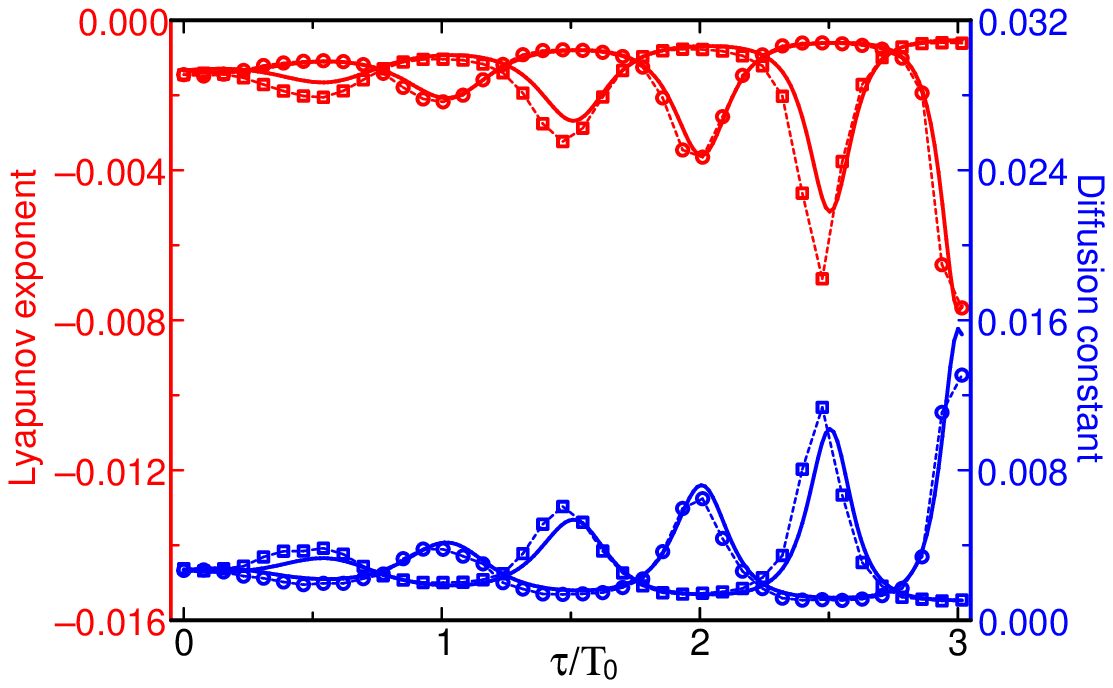}\\[5pt]
(b)&\hspace{-6mm}\includegraphics[width=0.435\textwidth]%
 {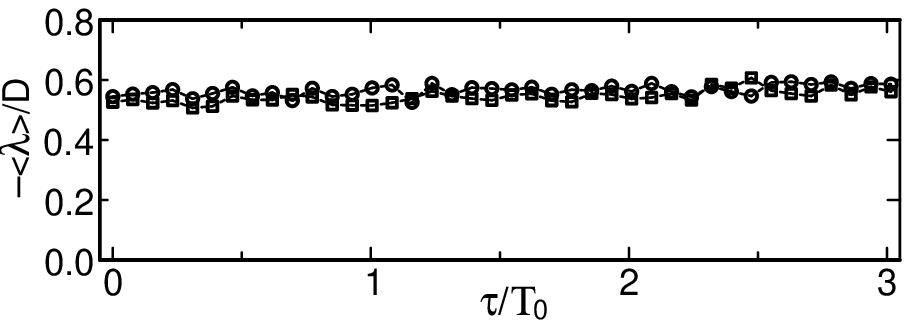}
\end{tabular}
}
  \caption{(Color online)
Dependencies of Lyapunov exponent $\la\lambda\ra$ (a: upper
graphs) and diffusion constant $D$ (a: lower graphs) on delay time
$\tau$ for the Van der Pol oscillator~(\ref{vdp}) with $\mu=0.7$
subject to white Gaussian noise of strength $\e^2=0.01$ and the
linear delayed feedback of strength $k=0.06$ (squares) and
$k=-0.06$ (circles). Oscillation period of the control-free
noiseless system $T_0\approx2\pi/0.96$\,. The solid lines present
analytical dependencies [Eqs.\,(\ref{e10}),(\ref{e11})]. (b): The
inconstancy of ratio $-\la\lambda\ra/D$ is not resolvable against
the background of the calculation inaccuracy.}
  \label{fig1}
\end{figure}

Fig.\,\ref{fig1} shows the effects of a linear delayed feedback on
the diffusion constant (DC) and the Lyapunov exponent (LE)
measuring exponential growth rate of perturbations in the
system~(\ref{vdp}). Noteworthy, not merely the LE and the DC are
crucially magnified or suppressed simultaneously at $\tau/T_0$
(here $T_0$ is the oscillation period of the control-free
noiseless system) being integers and half-integers, but even their
ratio remains nearly constant as $\tau$ changes (see
Fig.\,\ref{fig1}b).

Let us develop a phase description of the system. One can
parameterize the states of a limit cycle system on the limit cycle
by the oscillation phase $\ph$ uniformly growing in the course of
temporal evolution. Such an oscillator subject to weak noise and
feedback stays in the vicinity of this cycle, and its evolution
may be still described within the framework of the conventional
phase approximation~\cite{Kuramoto-2003}. Close to the bifurcation
point, {\it i.e.}, for $\mu\to0$, the Van der Pol oscillator has a
nearly circular limit cycle: $x_0=2\cos\ph$,
$\dot{x}_0=-2\sin\ph$, and the phase equation for the
system~(\ref{vdp}) reads
(cf.\,\cite{Goldobin-Rosenblum-Pikovsky-2003})
\be
\dot\ph=\Omega_0
 +a\,g[\ph(t-\tau),\ph(t)]+\e f[\ph(t)]\circ\xi(t)\,,
\ee{e01}
\noindent
where $\Omega_0=2\pi/T_0$ is the inherent cyclic frequency of the
system, $a=k/2$, $g=\sin[\ph(t-\tau)-\ph(t)]$, the sign
``$\circ$'' means a Stratonovich form of the equation,
$2\pi$-periodic function $f(\ph)$ is the sensitivity to noise. For
an additive noise as in Eq.\,(\ref{vdp}),
$f(\ph)=(2\Omega_0)^{-1}\cos\ph$
(cf.\,\cite{Goldobin-Rosenblum-Pikovsky-2003}), but we keep $f$
for generality. Noteworthy, in Ref.\,\cite{Masoller-2002},
Eq.\,(\ref{e01}) has been used to describe the evolution of the
phase of an optical field in a laser with a weak optical feedback.

For a small perturbation $\alpha$, one finds
\begin{eqnarray}
\dot\alpha=a\cos[\ph(t-\tau)-\ph(t)](\alpha(t-\tau)-\alpha(t))\nonumber\\
 {}+\e f'[\ph(t)]\,\alpha(t)\circ\xi(t)\hspace{-10mm}\nonumber
\end{eqnarray}
\noindent
(the prime stands for derivative with respect to the argument).
Therefore, the instant exponential growth rate $\lambda(t)$ obeys
\begin{eqnarray}
\lambda(t)=a\cos[\ph(t-\tau)-\ph(t)]
\big(e^{-\int_{t-\tau}^t\lambda(t_1)\mathrm{d}t_1}-1\big)\nonumber\\
 {}+\e f'[\ph(t)]\circ\xi(t)\,,\hspace{-2mm}
\label{e02}
\end{eqnarray}
\noindent
here we have made use of
$\alpha(t)\propto\exp(\int^t\lambda(t_1)\mathrm{d}t_1)$. Note,
that the LE is the mean value $\la\lambda\ra$.

\begin{figure}[!t]
{\sf
\begin{tabular}{lc}
(a)&\hspace{-6mm}\includegraphics[width=0.435\textwidth]%
 {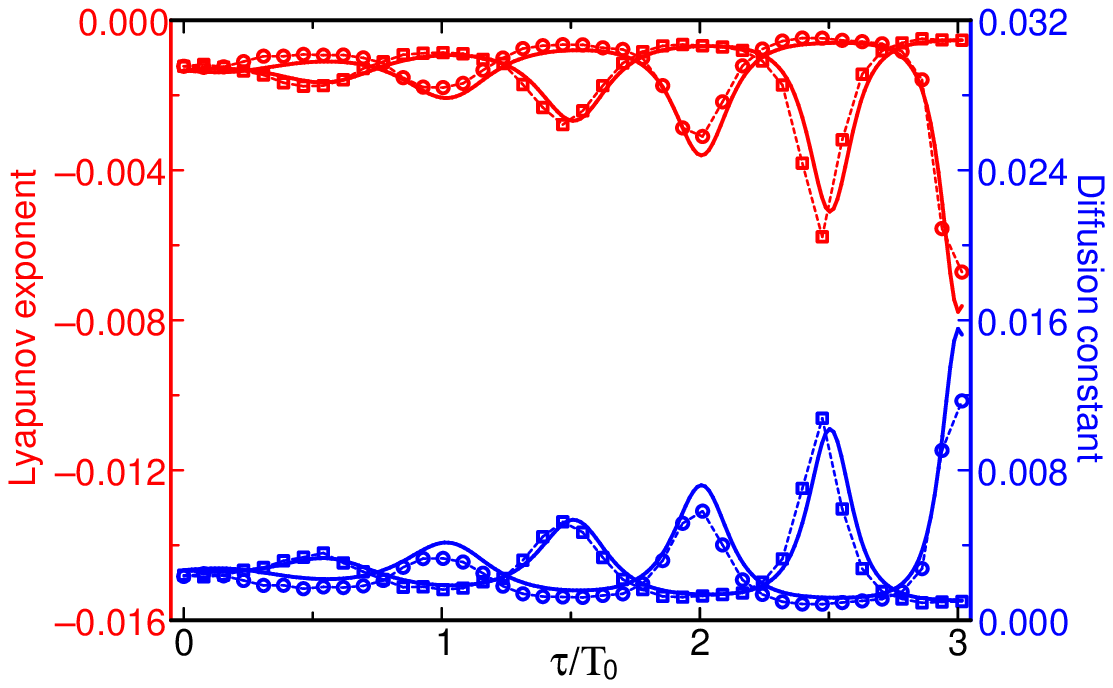}\\[5pt]
(b)&\hspace{-6mm}\includegraphics[width=0.435\textwidth]%
 {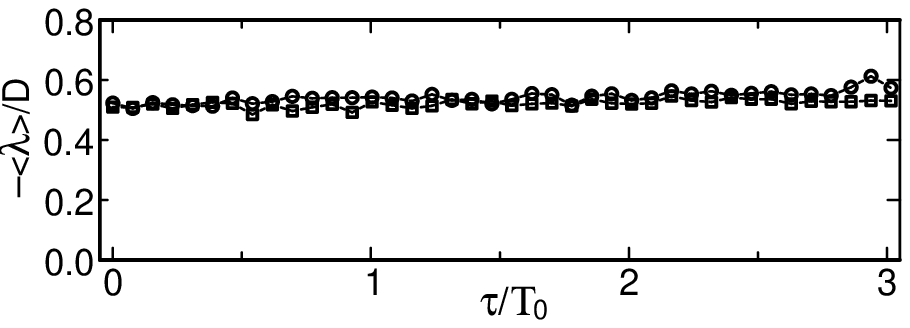}
\end{tabular}
}
  \caption{(Color online)
Same dependencies as in Fig.\,\ref{fig1} for the same parameter
values but for red Gaussian noise
$\zeta(t)=T^{-1}\int_{t-T}^t\xi(t_1)\,\mathrm{d}t_1$ with
$T=1.2$\,. For notation see caption to Fig.\,\ref{fig1}.}
  \label{fig2}
\end{figure}

For further analysis, it is more convenient to consider equations
in Ito form. Eqs.\,(\ref{e01}),(\ref{e02}) read
\begin{eqnarray}
\dot\ph=\Omega_0\!
 +a\sin[\ph(t\!-\!\tau)\!-\!\ph(t)]\!+\e^2f'f\!+\e f[\ph(t)]\xi(t),
\label{e03}\\
\lambda(t)=a\cos[\ph(t-\tau)-\ph(t)]
\big(e^{-\int_{t-\tau}^t\lambda(t_1)\,\mathrm{d}t_1}-1\big)\quad\;\nonumber\\
 {}+\e^2f''f+\e f'[\ph(t)]\,\xi(t).
\label{e04}
\end{eqnarray}
\noindent
The terms ahead of the noisy ones describe the Stratonovich drift.
Recall, in Ito form (with the Stratonovich drift included
explicitly) the instant value $\ph(t)$ is independent of the
instant value $\xi(t)$ taken at the same time moment $t$.

Let us explicitly introduce the mean frequency $\Omega$ and the
instant frequency deviation $v$; $\ph\equiv\Omega t+\psi$,
$\dot\psi=v$, $\la v\ra=0$. For a weak noise and a small feedback
strength ($\e\ll1$, $|a|\ll1$) the instant frequency fluctuations
are small ($v\ll1$), and Eqs.\,(\ref{e03}),(\ref{e04}) yield, up
to the main order of accuracy,
\begin{eqnarray}
&&\hspace{-5mm}
\Omega=\Omega_0-a\sin{\Omega\tau}\,,
\label{e05}\\
&&\hspace{-5mm}
 v=-a\cos{\Omega\tau}\,[\psi(t\!-\!\tau)\!-\!\psi(t)]+\e^2f'f+\e f\,\xi(t),
\label{e06}\\[3pt]
&&\hspace{-5mm}
\lambda=a\big(\cos{\Omega\tau}
 +\sin{\Omega\tau}\,[\psi(t-\tau)-\psi(t)]\big)
\nonumber\\
&&\hspace{ 3mm}
\times\big(e^{-\int_{t-\tau}^t\lambda(t_1)\,\mathrm{d}t_1}-1\big)
 +\e^2f''f+\e f'\,\xi(t)\,.\quad
\label{e07}
\end{eqnarray}

\begin{figure}[!t]
{\sf
\begin{tabular}{lc}
(a)&\hspace{-6mm}\includegraphics[width=0.435\textwidth]%
 {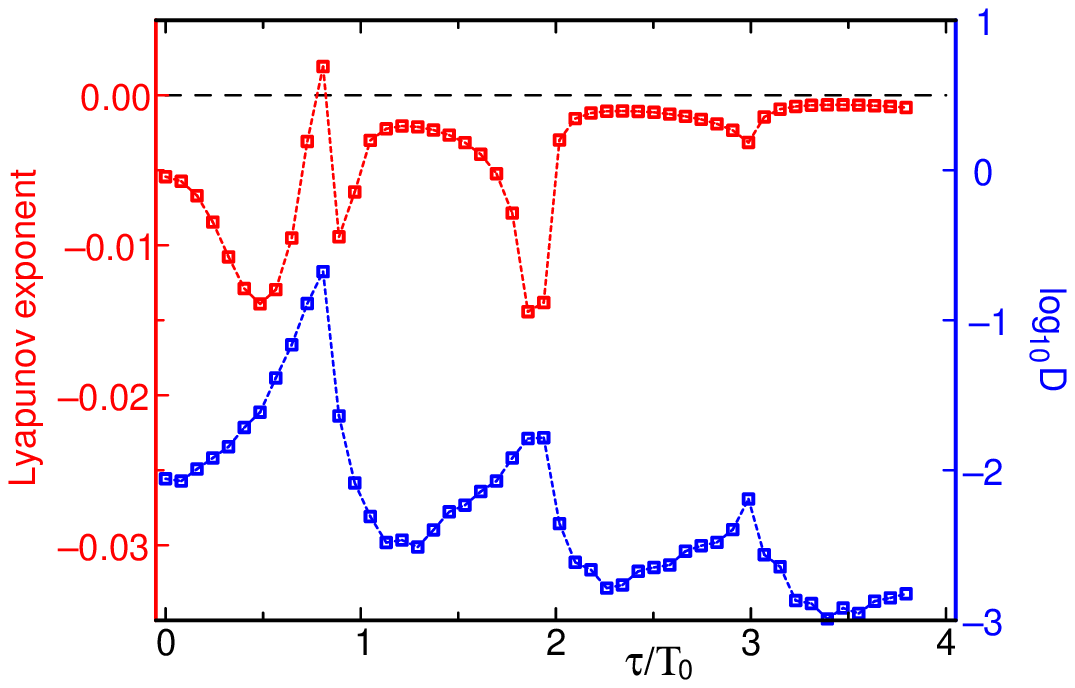}\\[5pt]
(b)&\hspace{-6mm}\includegraphics[width=0.435\textwidth]%
 {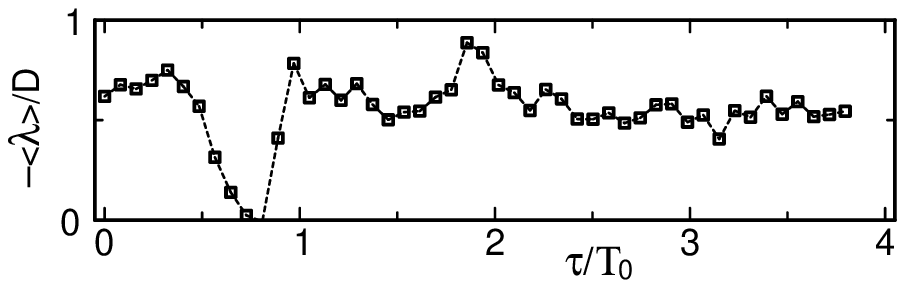}
\end{tabular}
}
  \caption{(Color online)
Dependencies $\la\lambda\ra(\tau)$ (a: upper graph) and $D(\tau)$
(a: lower graph) for the Van der Pol--Duffing
oscillator~(\ref{vdpd}) with $\mu=0.2$ ($T_0\approx2\pi/2.02$)
subject to white Gaussian noise, $\e=0.05$, and the delayed
feedback, $k=0.06$. For description and notation of (b), see
Fig.\,\ref{fig1} caption.
}
  \label{fig3}
\end{figure}

Let us now find the LE. Assuming $v$ and the fluctuating part
$\widetilde{\lambda}$ of $\lambda$, obeying
\be
\widetilde{\lambda}(t)\approx
 -a\cos{\Omega\tau}\int_{t-\tau}^t\widetilde{\lambda}(t_1)\,\mathrm{d}t_1
 +\e f'(\Omega t)\,\xi(t)\,,
\ee{e08}
\noindent
to be Gaussian, one can employ Furutsu--Novikov
formula~\cite{Furutsu-1963-Novikov-1964} to obtain from
Eq.\,(\ref{e07})~\footnote{Notice, Eq.(\ref{e07}) should not be
linearized with respect to $\lambda$.}
\be
\la\lambda\ra=a\cos{\Omega\tau}\Big(-\la\lambda\ra\tau
 +\frac{1}{2}\Big\la\big[\!\int_{t-\tau}^t\!\!\!\widetilde{\lambda}(t_1)\,\mathrm{d}t_1\big]^2\Big\ra\Big)
 -\e^2\la f'^2\ra_\ph
\ee{e09}
\noindent
(here $\la...\ra_\ph$ stands for an average over the phase $\ph$).
The value
$I\equiv\Big\la\big[\!\int_{t-\tau}^t\!\!\!\widetilde{\lambda}(t_1)\,\mathrm{d}t_1\big]^2\Big\ra$
can be evaluated from Eq.\,(\ref{e08})
 (similarly to $\la v(t_1)v(t_2)\ra$ in Ref.\,\cite{Goldobin-Rosenblum-Pikovsky-2003});
\begin{eqnarray}
&&\hspace{-3mm}
 I=\e^2\la f'^2\ra_\ph\frac{\tau}{\pi}\int_{-\infty}^{+\infty}
 \left|\frac{ix}{1-e^{-ix}}+a\tau\cos{\Omega\tau}\right|^{-2}\!\!\mathrm{d}x\nonumber\\[3pt]
&&\hspace{-2mm}
 =\e^2\la f'^2\ra_\ph\left\{
 \begin{array}{cc}
 \displaystyle
 \frac{2\tau}{1+a\tau\cos{\Omega\tau}}\,,
 &
 \mbox{\small for }\,a\tau\cos{\Omega\tau}\!>\!-1;\\[8pt]
 \displaystyle
 \!\!\frac{2\tau}{\pi}\!\left[\frac{2}{1+a\tau\cos{\Omega\tau}}\right]^{\frac{8}{7}}\!\!,
 &
 \mbox{\small for }\,a\tau\cos{\Omega\tau}\!<\!-1.
 \end{array}
 \right.\nonumber
\end{eqnarray}
\noindent
For $a\tau>1$, Eq.\,(\ref{e05}) exhibits multistability of mean
frequency
$\Omega$~\cite{Masoller-2002,Niebur_etal-1991,Goldobin-Rosenblum-Pikovsky-2003},
which results in the violation of the basic assumptions of our
analytical theory. Hence, the case $a\tau\cos{\Omega\tau}\!<\!-1$
may be ignored as meaningless, and after substitution of $I$,
Eq.\,(\ref{e09}) reads
\be
\la\lambda\ra=
-\frac{\e^2\la f'^2\ra_\ph}{(1+a\tau\cos{\Omega\tau})^2}\,,
\ee{e10}
\noindent
whereas the DC has been already evaluated in
Ref.\,\cite{Goldobin-Rosenblum-Pikovsky-2003};
\be
D=\frac{2\,\e^2\la f^2\ra_\ph}{(1+a\tau\cos{\Omega\tau})^2}\,.
\ee{e11}
\noindent
Therefore,
\be
 -\frac{\la\lambda\ra}{D}=\frac{\la f'^2\ra_\ph}{2\,\la f^2\ra_\ph}
 =const
\ee{e12}
\noindent
which is $1/2$ for $f(\ph)=(2\Omega_0)^{-1}\cos{\ph}$ as for the
Van der Pol system in Fig.\,\ref{fig1}. Note that, due to the
deformation of the limit cycle at $\mu=0.7$, relation~(\ref{e12})
is more accurate than Eqs.\,(\ref{e10}),(\ref{e11}) where the term
$a\tau\cos{\Omega\tau}$ is specific to
$g=\sin[\ph(t-\tau)-\ph(t)]$ (see Fig.\,\ref{fig1} where
$-\la\lambda\ra/D\approx0.55$).

While deriving Eqs.\,(\ref{e10}),(\ref{e11}) we nowhere utilized
that the noise is $\delta$-correlated. Remarkably, the results
remain valid for colored noise, {\it e.g.}, a red one (see
Fig.\,\ref{fig2}), results for which coincide with the one for
white noise almost up to the numerical calculation inaccuracy and
$-\la\lambda_\mathrm{red}\ra/D_\mathrm{red}\approx0.53$.

\begin{figure}[!t]
\center{\includegraphics[width=0.435\textwidth]%
 {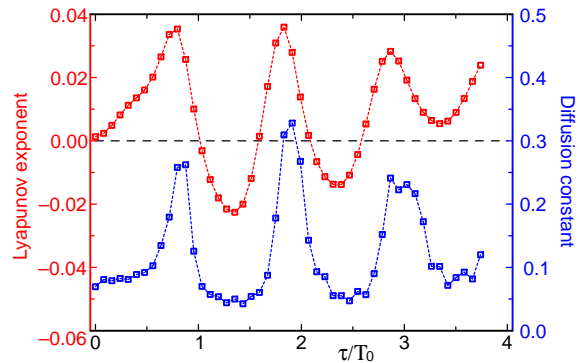}
}
  \caption{(Color online)
Same dependencies as in Fig.\,\ref{fig3}a for the same Van der
Pol--Duffing system and feedback strength but a larger noise,
$\e=0.374$\,.}
  \label{fig4}
\end{figure}

\begin{figure}[!b]
\center{\includegraphics[width=0.475\textwidth]%
 {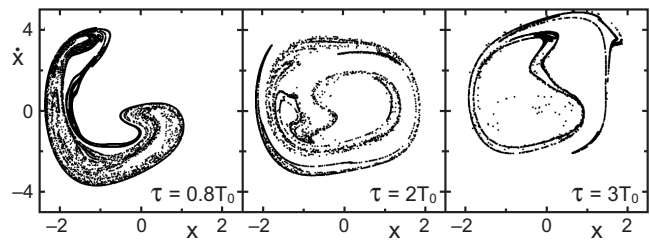}
}
  \caption{The snapshots
of the ensemble of $20\,000$ identical Van der Pol--Duffing
oscillators ($\mu=0.2$) driven by common white Gaussian noise
($\e=0.374$) have a fine structure for $\tau$ indicated in the
plots (cf.\ to the LE in Fig.\,\ref{fig4}).}
  \label{fig5}
\end{figure}

For a strong noise the phase description always leading to a
negative LE is not applicable, and even positive LEs have been
reported~\cite{Pikovsky-1984,Goldobin-Pikovsky-2005b,telegraph_noise,impulse_noise,Goldobin-Pikovsky-2006}.
This case can be treated only numerically. For this sake, a
simulation for the Van der Pol--Duffing oscillator
\be
\ddot{x}-\mu(1-x^2)\dot{x}+x+x^3=k[\dot{x}(t-\tau)-\dot{x}(t)]+\e\,\xi(t)
\ee{vdpd}
\noindent
exhibiting positive LEs for a moderate
noise~\cite{Goldobin-Pikovsky-2005b} has been performed. Let us
note that for a non-large noise (Fig.\,\ref{fig3}), ratio
$-\la\lambda\ra/D$ is changed not greater than by $20\%$, while
the DC and the LE are changed by factor $\approx20$, in a broad
range of $\tau$ (the only exception is interval $[T_0/2,T_0]$ near
the domain where the LE is positive).

At $\e=0.374$ (Fig.\,\ref{fig4}), where the control-free Van der
Pol--Duffing oscillator just becomes unstable (unreliable),
$\la\lambda\ra|_{\tau=0}\approx0$, the linear delayed feedback
leads to maximal positive LEs for integer $\tau/T_0$ and minimal
(not always negative) LEs for half-integer $\tau/T_0$. Concerning
the interpretation of dependence $\la\lambda\ra(\tau)$, note the
following. In the absence of the feedback control, an
intermittency of epochs of positive and negative local LEs
(``local'' means evaluated over a finite time interval) takes
place and the transition to positive LE is related to a plain
quantitative prevalence of the former over the latter
(cf.\,\cite{Goldobin-Pikovsky-2006}). The feedback affects
(magnifies or suppresses) the local LEs over these epochs
non-uniformly, thus shifting balance between these epoches and
bringing about a domination of positive local LEs for integer
$\tau/T_0$ and negative ones for half-integer $\tau/T_0$. For
positive ``global'' LEs, phase diffusion owes mainly not to
stochasticity but to chaos (samples of snapshot chaotic
attractors, see Ref.\,\cite{Yu-Chen-Ott-1991}, are presented in
Fig.\,\ref{fig5}). As a result, here the DC is diminished where
the LE is minimal.


Summarizing, for a weak white or colored Gaussian
noise~\footnote{Perhaps, the diffusionless noises as a blue one
should be excluded because for them, unlikely to the case studied,
phase diffusion would be purely due to an interaction between the
noise and the system nonlinearity.}, highly stable response
(reliability) to a noisy driving is observed when phase diffusion
is strong ({\it i.e.}, the coherence is weak). {\it Vice versa},
for small diffusion ({\it i.e.}, highly coherent oscillations)
response is weakly stable [Figs.\,1,2, Eq.\,(\ref{e12})]. In
particular, this imposes strong limitations on the implementation
of the technique of coherence improvement by virtue of a linear
delayed feedback. For instance, in an ensemble of uncoupled
identical self-sustained oscillators synchronized by a common
external noisy driving, small intrinsic noise is always present
and leads to spreading of oscillator phases:
$\Delta\ph\propto\e_\mathrm{in}/\sqrt{-\la\lambda\ra}$
($\e_\mathrm{in}$ is the amplitude of intrinsic noise,
cf.\,\cite{Goldobin-Pikovsky-2005b}). In such an ensemble the
delayed feedback improvement of the coherence results in a mutual
spreading of oscillator phases which may be sometimes undesirable.

For a strong noise being capable to create a positive Lyapunov
exponent, {\it i.e.}, antireliability, chaotic contribution to
phase diffusion may prevail over the stochastic one, and then an
enhanced coherence occurs for the maximal reliability
(Fig.\,\ref{fig4}).

Detailed calculation of the Lyapunov exponent in the above text
serves the purpose to disclose the essentially different nature of
various contributions to the Lyapunov exponent and the diffusion
constant. However, the final quantitative effect of delayed
feedback on these dissimilar properties of oscillatory systems
somewhat surprisingly turns out to be identical. The reported
phenomenon being valid for a general class of limit cycle
oscillators is, thus, neither intuitively expected nor trivial.

I thank A.\ Pikovsky and M.\ Zaks for useful discussions and
appreciate financial support from the Foundation ``Perm
Hydrodynamics,'' CRDF (Grant No.\ Y5–P–09–01), and MESRF (Grant
No.\ 2.2.2.3.16038).


\end{document}